\documentclass[]{rmaa}

\usepackage{paralist}
\usepackage[T1]{fontenc}

\usepackage{psfrag,color}

\usepackage[latin1]{inputenc}
\usepackage{epsfig}
\usepackage{amsmath}
\usepackage{graphicx}
\usepackage{subfig}
\usepackage{multirow}
\usepackage{array}
\usepackage{rotating}
\usepackage[normalem]{ulem}

\def\pnv{V1838~Aql}

\title{Extensive photometry of V1838 Aql during the 2013 superoutburst} 

\author{
J. Echevarr\'ia\altaffilmark{1},
E. de Miguel\altaffilmark{2},
J. V. Hern\'andez~Santisteban\altaffilmark{3},
R. Michel\altaffilmark{4}, \\
R. Costero\altaffilmark{1},
L. J. S\'anchez\altaffilmark{1},
A. Ruelas-Mayorga\altaffilmark{1},
J. Olivares\altaffilmark{5},\\
D. Gonz\'alez-Buitrago\altaffilmark{6}, 
J.L. Jones\altaffilmark{7},
A. Oskanen\altaffilmark{8},
W. Goff\altaffilmark{9},
J. Ulowetz\altaffilmark{10},\\
G. Bolt\altaffilmark{11},
R. Sabo\altaffilmark{12},
F.-J Hambsch\altaffilmark{13},
D. Slauson\altaffilmark{14},
and W. Stein\altaffilmark{15}}

\altaffiltext{1}{
Instituto de Astronom\'ia, 
Universidad Nacional Aut\'onoma de M\'exico, 
Apartado Postal 70-264, Ciudad Universitaria, 
M\'exico D.F., C.P. 04510, M\'exico}
\altaffiltext{2} {Departamento de Ciencias Integradas, Universidad de Huelva, E--21071 Huelva, Spain}
\altaffiltext{3}{
Anton Pannekoek Institute for Astronomy, University of Amsterdam, Science Park 904, 1098XH Amsterdam, the Netherlands}
\altaffiltext{4}{Instituto de Astronom\'ia, 
Universidad Nacional Aut\'onoma de M\'exico, 
Apartado Postal 877, 
Ensenada, Baja California, C.P. 22830 M\'exico}
\altaffiltext{5}{Laboratoire d'Astrophysique de Bordeaux, Univ. Bordeaux, CNRS, B18N, all\'ee Geoffroy Saint-Hilaire, 33615 Pessac, France}
\altaffiltext{6}{Departament of Physics and Astronomy 4129, Frederick Reines Hall, University of California, Irvine, CA, 92697-4575}
\altaffiltext{7}{
CBA--Oregon, Jack Jones Observatory, 22665 Bents Road NE, Aurora, OR, USA}
\altaffiltext{8}{
CBA--Finland, Hankasalmi Observatory, Verkkoniementie 30, FI--40950 Muurame, Finland}
\altaffiltext{9}{
CBA--California, 13508 Monitor Lane, Sutter Creek, CA 95685, USA}
\altaffiltext{10}{
CBA--Illinois, Northbrook Meadow Observatory, 855 Fair Lane, Northbrook, IL 60062, USA}
\altaffiltext{11}{
CBA--Australia, 295 Camberwarra Drive, Craigie, Western Australia 6025}
\altaffiltext{12}{
CBA--Montana, 2336 Tailcrest Dr., Bozeman, MT 59718, USA}
\altaffiltext{13}{
CBA--Mol, Andromeda Observatory, Oude Bleken 12, B--2400 Mol, Belgium}
\altaffiltext{14}{
CBA--Iowa, Owl Ridge Observatory, 73 Summit Avenue NE, Swisher, IA 52338, USA}
\altaffiltext{15}{
CBA--Las Cruces, 6025 Calle Paraiso, Las Cruces, NM 88012, USA}

\shortauthor{J. Echevarr\'ia et al.}
\shorttitle{Photometry of V1838 Aql during superoutburst}

\abstract{

We present an in-depth photometric study of the 2013 superoutburst of the recently discovered cataclysmic variable V1838 Aql and subsequent photometry near its quiescent state. A careful examination of the development of the superhumps is presented. Our best determination of the orbital period is $P_{\rm{orb}} = 0.05698(9)$  days, based on the periodicity of early superhumps.  Comparing the superhump periods at stages A and B with the early superhump value we derive  a period excess of $\epsilon = 0.024(2)$ and a mass ratio of $q = 0.10(1)$. We suggest that \pnv\  is approaching the orbital period minimum and thus has a low-mass star as a donor instead of a sub-stellar object.}

\resumen{

Presentamos un estudio fotom\'etrico detallado de la super-erupci\'on de \pnv , una variable catacl\'ismica recientemente descubierta, desde el m\'aximo en 2013 hasta su regreso al m\'inimo. Examinamos en detalle la evoluci\'on de los \textit{superhumps}.
Determinamos el per\'iodo orbital $P_{\rm{orb}} = 0.05698(9)$  d a partir de la periodicidad de los \textit{superhumps} tempranos. Comparando los per\'iodos de \textit{superhumps} en las etapas A y B con el valor del per\'iodo orbital, derivamos un valor del cambio en el per\'iodo orbital de  $\epsilon = 0.024(2)$ y un cociente de masa para el sistema de $q = 0.10(1)$. Sugerimos que \pnv\ se est\'a acercando al m\'inimo per\'iodo orbital, por lo que la secundaria ser\'ia una estrella de baja masa y no un objeto sub-estelar.}

\addkeyword{stars: cataclysmic variables, dwarf novae, individual (V1838 Aql) -- techniques: photometric}
\begin{document}
\maketitle

\section{Introduction}
\label{sec:intro}

Cataclysmic variables (CVs) are close binary systems where a white dwarf (WD) accretes from a low-mass star via Roche-lobe overflow, often creating an accretion disc \citep[for a review see][]{war95}. A large fraction of CVs belong to the subclass of dwarf novae (DNe). They undergo recurrent outbursts with typical amplitudes of $\sim$2--6 mag in the optical, which are commonly accepted to be caused by a thermal-viscous instability in the disc \citep{osa74}. In addition, SU~UMa-type DNe (subclass of DNe systems with short orbital periods, $P_{\rm orb} < 2.5$ hr) exhibit occasional eruptions that are less frequent, longer lasting, and slightly brighter (by $\sim$ 0.5--1.0 mag) than the normal outbursts. The key feature during these so-called superoutbursts is the presence of superhumps a modulation in the light curve with an underlying periodicity, $P_{\rm sh}$, a few percent longer than the orbital period. They are thought to arise from a precessing non-axisymmetric disc \citep{vog82}, with the eccentricity being produced by the tidal instability developed at the radius of the 3:1 resonance \citep[]{whi88}.
The analysis of the timing and evolution of such light oscillations provides estimates of the system's parameters using empirical \citep{pat05} and theoretical \citep{kat13} relationships between the superhump/orbital period excess, $\epsilon \equiv (P_{\rm sh} - P_{\rm orb})/ P_{\rm orb}$, and the binary mass ratio $q\equiv M_2/M_1$. Several 
$\epsilon(q)$ relations have been proposed by these authors based on different stages of the superhumps, although \citet{otu16} show that the scatter in the $\epsilon$-$q$ diagram is considerable for very short orbital periods. This is due to a lack of objects with dynamically confirmed small values of $q$ \citep{pat11,kat13}. 

Among the SU UMa--type systems, there is a large group that accumulates around 
the minimum of the orbital period distribution of CVs 
($P_{\rm orb}\sim78$ min) \citep{pas81,gan09,kni11}. These are binaries with extremely low mass-transfer rates, named WZ~Sge-type objects, which are characterised by rare (commonly detected every $\sim$ 10 years), and large amplitude superoutbursts of duration of $\sim$ 30 days, caused by an instability in low viscosity accretion discs with $\alpha\sim0.01--0.001$ \citep{sma93,osa94}. Some of them are systems currently evolving towards longer periods and are collectively known as period bouncers \citep[e.g.][]{ldm03}. These binaries are expected to harbour a sub-stellar secondary companion \citep{how97}, i.e. brown dwarfs \citep[e.g.][]{lea06,har16,her16,neu17}. 

Worth noting is that WZ Sge-type objects are characterised not only by a long superoutburst recurrence time in comparison to typical SU UMa stars, but also by the presence of early superhumps (double-wave modulation) during the first few days of the eruption, with a periodicity ($P_{\rm esh}$) essentially equal to the orbital period of the binary \citep[details in][]{oea91,kat15}.
The advent of all-sky surveys \citep[e.g.][]{bre14} and worldwide citizen-telescope networks have contributed to the discovery of a large population of  faint DNe. Among these discoveries, the elusive population of short-period systems, in particular period bouncers, has been found and investigated  \citep{pat11,cop16,otu16}. 

The discovery of a new transient, initially proposed as possible nova, was reported by Itagaki on 2013 May 31. Henden\footnote{\url{http://ooruri.kusastro.kyoto-u.ac.jp/mailarchive/vsnet-alert/15772)}} pointed out that the colour indices of the object and the un-reddened field suggested a DN rather than a nova. As pointed out by Hurst\footnote{\url{www.theastronomer.org/tacirc/2013/e2919.txt}}, Kojima reported a pre-discovery image on 2013 May 30.721 UT, when the magnitude was at about 9.8 mag (un-filtered). A CBET report of this new DNe in Aquila, can be found in \citet{ita13}.


Although a preliminary analysis of the behaviour of V1838 Aql (originally designated as PNV J19150199+0719471) was published by \citet{kato14}, we present here a full analysis of the superhump behaviour based on our extensive data.

In section \ref{sec:observations} we present the observations and their reduction methods. The photometric data and the period analysis are presented in section \ref{sec:images}, while in section \ref{sec:discus} we address the discussion of our results. We present our conclusions in section \ref{sec:conclusions}.

\section{Observations and Reduction}
\label{sec:observations}

Photometric observations in the $V$ band were obtained in 2013 during the nights of June 3, 4, 5, 6, 17, 18, 28 and September 2 and 25 at the 0.84 m telescope of the Observatorio Astron\'omico Nacional at San Pedro M\'artir (SPM). We used the Blue-ESOPO CCD detector\footnote{\url{http://www.astrossp.unam.mx/indexspm.html}} on a 2$\times$2 binning configuration. The exposure time of the SPM observations varied between 10 and 30~s. In addition, time-series photometry of the superoutburst were obtained from 10 observatories of the Center for Backyard Astrophysics (CBA) --a network of small (0.2--0.4~m) telescopes that covers a wide range in terrestrial longitude.  \citet{skillman93} and \citet{demiguel16} describe the methods and observing stations of the CBA network. These observations amounted to 162 separate time-series during 58 nights from June 1 
to August 2, 2013, and the typical exposure time ranged from 20 to 120 seconds depending on the brightness of \pnv. Nearly half of these observations were obtained in $V$ light,  while the rest (mainly during the post-outburst regime) were unfiltered. We did not attempt any absolute calibration of the data during the eruption, but the magnitude scale is expected to resemble closely $V$ magnitudes with a zero-point uncertainty of $\sim 0.05$ mag. Further observations in the $R$ band using the 2.1~m telescope at SPM during 2018, July 18 were conducted. Unfortunately, the weather was unstable and we only managed to obtain differential photometry over three orbital cycles. In the following, we report times and refer to specific dates in a truncated form defined as $\text{HJD}- 2,456,000$.

\section{Photometry and period analysis}\label{sec:images}





\subsection{Photometric observations}
\label{sec:global_lc}
 
Most of our photometric observations come from the CBA network, with additional $V$-band observations obtained with the 0.84 m telescope at SPM at some critical stages of the outburst and during the late decline.\footnote{All CBA and SPM data are available on request.} We observed the typical pattern seen in SU~UMa stars during superoutburst: a plateau phase --lasting $\sim$25 days, from HJD 444 to 469-- where the mean brightness varies smoothly from 10.5 to 13.0 mag, followed by a rapid decline ($\sim$3 mag in 2 days) at the end of the main eruption. The subsequent fading towards quiescence was at a rather low rate ($\sim$0.035 mag d$^{-1}$), and even 3.5 years after the end of the eruption, the system was found $\sim$0.5 mag above the pre-eruption quiescent brightness. However, based on the observations obtained with the NTT telescope (La Palma, Spain), we confirmed that the object had reached the pre-outburst level by June 2017. These observations and the general spectral distribution at quiescence have no further relevance here and will be discussed in a future publication. 


 \subsection{Full analysis of the different stages of the superhump}


Our primary tool for studying periodic signals was the Period 04 package \citep{period04}. First, we subtracted the mean and (linear) trend from each individual light curve and formed nightly-spliced light curves. Then, after combining light curves from adjacent nights, a search for periodic signals was done. This approach allows to improve the frequency resolution, but it has to be implemented with caution, since variations in the amplitude and/or the period of the modulation --both effects known to afflict erupting DNe-- can distort the outcome of the frequency analysis.


A general view of the superhump transitions can be looked up by identifying the different stages of the superhumps: early superhumps, {\it stages} A, B and C as well as the post-outburst stage \citep[see][for this terminology in our general discussion]{kato09,kato14}. Thus, we looked at the time variations of the superhump period and its amplitude by examining the variation in time with respect to a well-defined feature of the superhump signal. 

After this general analysis, a detailed explanation of these stages was made. First, we derived the timings of superhump maxima. A total of 310 times of superhump maxima were identified in the light curves in the interval HJD~449.6--498.6 days. These maxima are shown in Table~\ref{Table1}. The early stage of the eruption was not considered, since the signal there was of very low amplitude and individual maxima were not well defined. A linear regression to these timings provides the following test ephemeris:

\begin{equation}
T_{\rm max} ({\rm HJD}) = 2,456,452.8035(23) + 
0.058191(6) \, E \,   .
\label{ephem_all}
\end{equation}

\begin{table*}[!t]\centering
\setlength{\tabnotewidth}{\columnwidth}\tablecols{12}
  \setlength{\tabcolsep}{0.5\tabcolsep}
\caption{Times of superhump maxima of \pnv \, during the 2013 superoutburst.\tabnotemark{*}}\label{Table1} 
\begin{footnotesize}
\begin{tabular}{cccccccccccc} 
\toprule
E\tabnotemark{**} & T\_max\tabnotemark{***} & O-C\tabnotemark{****} & E\tabnotemark{**} & T\_max\tabnotemark{***} & O-C\tabnotemark{****} & E\tabnotemark{**} & T\_max\tabnotemark{***} & O-C{****} & E\tabnotemark{**} & T\_max\tabnotemark{***} & O-C\tabnotemark{****} \\
\cmidrule(lr){1-3}\cmidrule(lr){4-6}\cmidrule(lr){7-9}\cmidrule(lr){10-12}
-51 & 449.7831 & -0.906 & 116 & 459.5372 & -0.283 & 189 & 463.8032 & 0.028  & 345 & 472.9088 & 0.506 \\
-50 & 449.8282 & -1.131 & 116 & 459.5365 & -0.295 & 190 & 463.8598 & -0.001 & 346 & 472.9681 & 0.524 \\
-50 & 449.8354 & -1.007 & 117 & 459.5944 & -0.301 & 190 & 463.8607 & 0.016  & 359 & 473.7247 & 0.527 \\
-49 & 449.8934 & -1.011 & 117 & 459.5939 & -0.309 & 195 & 464.1540 & 0.056  & 360 & 473.7820 & 0.511 \\
-49 & 449.8956 & -0.972 & 118 & 459.6513 & -0.322 & 196 & 464.2150 & 0.104  & 361 & 473.8381 & 0.474 \\
-34 & 450.7936 & -0.541 & 120 & 459.7703 & -0.278 & 197 & 464.2728 & 0.097  & 362 & 473.8938 & 0.431 \\
-33 & 450.8491 & -0.586 & 121 & 459.8289 & -0.271 & 198 & 464.3288 & 0.060  & 363 & 473.9561 & 0.502 \\
-22 & 451.4995 & -0.410 & 122 & 459.8842 & -0.320 & 199 & 464.3889 & 0.092  & 371 & 474.4228 & 0.523 \\
-21 & 451.5535 & -0.481 & 123 & 459.9432 & -0.307 & 200 & 464.4455 & 0.065  & 372 & 474.4782 & 0.475 \\
0   & 452.7934 & -0.174 & 131 & 460.4149 & -0.200 & 205 & 464.7456 & 0.222  & 373 & 474.5387 & 0.514 \\
1   & 452.8512 & -0.181 & 132 & 460.4708 & -0.239 & 206 & 464.7997 & 0.151  & 374 & 474.5940 & 0.464 \\
16  & 453.7190 & -0.268 & 132 & 460.4718 & -0.223 & 207 & 464.8537 & 0.080  & 375 & 474.6515 & 0.453 \\
17  & 453.7782 & -0.250 & 133 & 460.5281 & -0.254 & 212 & 465.1482 & 0.141  & 376 & 474.7119 & 0.490 \\
18  & 453.8358 & -0.261 & 133 & 460.5286 & -0.247 & 213 & 465.2072 & 0.154  & 377 & 474.7702 & 0.492 \\
19  & 453.8959 & -0.228 & 134 & 460.5883 & -0.220 & 214 & 465.2684 & 0.205  & 377 & 474.7705 & 0.497 \\
47  & 455.5203 & -0.313 & 134 & 460.5883 & -0.220 & 215 & 465.3263 & 0.201  & 377 & 474.7706 & 0.500 \\
48  & 455.5766 & -0.345 & 135 & 460.6459 & -0.231 & 234 & 466.4437 & 0.403  & 378 & 474.8263 & 0.456 \\
49  & 455.6356 & -0.331 & 136 & 460.7049 & -0.218 & 235 & 466.5055 & 0.465  & 378 & 474.8280 & 0.486 \\
50  & 455.6938 & -0.331 & 137 & 460.7625 & -0.227 & 236 & 466.5650 & 0.488  & 379 & 474.8868 & 0.497 \\
51  & 455.7512 & -0.345 & 138 & 460.8201 & -0.237 & 237 & 466.6212 & 0.453  & 379 & 474.8862 & 0.486 \\
52  & 455.8099 & -0.336 & 139 & 460.8780 & -0.243 & 239 & 466.7401 & 0.497  & 379 & 474.8909 & 0.566 \\
53  & 455.8690 & -0.321 & 139 & 460.8787 & -0.230 & 252 & 467.4850 & 0.298  & 380 & 474.9493 & 0.570 \\
54  & 455.9259 & -0.342 & 139 & 460.8792 & -0.221 & 253 & 467.5438 & 0.309  & 393 & 475.6979 & 0.435 \\
64  & 456.5073 & -0.351 & 154 & 461.7554 & -0.164 & 254 & 467.6029 & 0.324  & 394 & 475.7561 & 0.435 \\
65  & 456.5671 & -0.323 & 154 & 461.7576 & -0.126 & 255 & 467.6619 & 0.338  & 394 & 475.7527 & 0.376 \\
66  & 456.6232 & -0.361 & 155 & 461.8140 & -0.158 & 269 & 468.4730 & 0.277  & 395 & 475.8151 & 0.449 \\
69  & 456.7989 & -0.341 & 156 & 461.8764 & -0.085 & 271 & 468.5903 & 0.292  & 395 & 475.8118 & 0.392 \\
70  & 456.8561 & -0.358 & 156 & 461.8756 & -0.098 & 272 & 468.6448 & 0.229  & 395 & 475.8140 & 0.430 \\
81  & 457.4979 & -0.328 & 157 & 461.9311 & -0.144 & 285 & 469.4134 & 0.436  & 396 & 475.8725 & 0.436 \\
82  & 457.5560 & -0.330 & 167 & 462.5166 & -0.083 & 286 & 469.4648 & 0.320  & 396 & 475.8737 & 0.456 \\
83  & 457.6146 & -0.323 & 168 & 462.5755 & -0.070 & 287 & 469.5288 & 0.420  & 397 & 475.9309 & 0.439 \\
84  & 457.6733 & -0.314 & 169 & 462.6334 & -0.076 & 288 & 469.5973 & 0.598  & 406 & 476.4504 & 0.366 \\
85  & 457.7303 & -0.335 & 170 & 462.6916 & -0.075 & 289 & 469.6537 & 0.567  & 407 & 476.5093 & 0.380 \\
86  & 457.7893 & -0.320 & 171 & 462.7489 & -0.090 & 290 & 469.7051 & 0.451  & 411 & 476.7418 & 0.374 \\
87  & 457.8469 & -0.332 & 171 & 462.7500 & -0.072 & 291 & 469.7620 & 0.428  & 412 & 476.7950 & 0.288 \\
88  & 457.9044 & -0.343 & 171 & 462.7514 & -0.047 & 292 & 469.8186 & 0.400  & 413 & 476.8565 & 0.345 \\
92  & 458.1365 & -0.354 & 172 & 462.8101 & -0.039 & 292 & 469.8215 & 0.450  & 414 & 476.9143 & 0.338 \\
93  & 458.1941 & -0.365 & 172 & 462.8074 & -0.086 & 293 & 469.8789 & 0.437  & 415 & 476.9763 & 0.404 \\
97  & 458.4284 & -0.338 & 172 & 462.8078 & -0.078 & 303 & 470.4617 & 0.452  & 423 & 477.4346 & 0.280 \\
99  & 458.5459 & -0.318 & 173 & 462.8644 & -0.106 & 305 & 470.5709 & 0.329  & 425 & 477.5504 & 0.270 \\
102 & 458.7208 & -0.313 & 173 & 462.8635 & -0.122 & 306 & 470.6391 & 0.501  & 426 & 477.6130 & 0.346 \\
103 & 458.7807 & -0.283 & 178 & 463.1596 & -0.033 & 308 & 470.7548 & 0.489  & 427 & 477.6698 & 0.321 \\
104 & 458.8372 & -0.313 & 179 & 463.2169 & -0.049 & 309 & 470.8143 & 0.510  & 429 & 477.7832 & 0.270 \\
105 & 458.8942 & -0.333 & 180 & 463.2767 & -0.021 & 310 & 470.8750 & 0.554  & 429 & 477.7877 & 0.348 \\
106 & 458.9522 & -0.336 & 181 & 463.3344 & -0.030 & 325 & 471.7415 & 0.446  & 430 & 477.8417 & 0.275 \\
110 & 459.1872 & -0.298 & 182 & 463.3946 & 0.005  & 326 & 471.7985 & 0.424  & 430 & 477.8434 & 0.305 \\
111 & 459.2441 & -0.321 & 184 & 463.5105 & -0.003 & 334 & 472.2435 & 0.072  & 431 & 477.9022 & 0.315 \\
112 & 459.3025 & -0.316 & 185 & 463.5680 & -0.015 & 335 & 472.3005 & 0.051  & 432 & 477.9570 & 0.257 \\
113 & 459.3617 & -0.300 & 186 & 463.6273 & 0.003  & 342 & 472.7271 & 0.381  & 435 & 478.1406 & 0.412 \\
115 & 459.4778 & -0.305 & 187 & 463.6863 & 0.019  & 343 & 472.7931 & 0.516  & 436 & 478.1927 & 0.308 \\
115 & 459.4767 & -0.323 & 188 & 463.7462 & 0.047  & 344 & 472.8467 & 0.437  & 437 & 478.2500 & 0.293 \\        
\toprule
\end{tabular}  
\end{footnotesize}
\end{table*}

\begin{table*}[!t]\centering
\setlength{\tabnotewidth}{\columnwidth}\tablecols{12}  
  \setlength{\tabcolsep}{0.5\tabcolsep}
TABLE 1. CONTINUED\\
\vspace{0.1cm} 
\begin{footnotesize}
\begin{tabular}{cccccccccccc} 
\toprule
E\tabnotemark{**} & T\_max\tabnotemark{***} & O-C\tabnotemark{****} & E\tabnotemark{**} & T\_max\tabnotemark{***} & O-C\tabnotemark{****} & E\tabnotemark{**} & T\_max\tabnotemark{***} & O-C{****} & E\tabnotemark{**} & T\_max\tabnotemark{***} & O-C\tabnotemark{****} \\
\cmidrule(lr){1-3}\cmidrule(lr){4-6}\cmidrule(lr){7-9}\cmidrule(lr){10-12}
438 & 478.3092 & 0.310 & 482 & 480.8586 & 0.120  & 535 & 483.9315 & -0.073 & 631 & 489.4996 & -0.386 \\
440 & 478.4235 & 0.273 & 483 & 480.9154 & 0.097  & 549 & 484.7405 & -0.170 & 632 & 489.5580 & -0.382 \\
441 & 478.4880 & 0.381 & 492 & 481.4423 & 0.151  & 550 & 484.8051 & -0.060 & 633 & 489.6199 & -0.318 \\
442 & 478.5401 & 0.277 & 493 & 481.5009 & 0.159  & 551 & 484.8538 & -0.222 & 647 & 490.4349 & -0.313 \\
443 & 478.6004 & 0.313 & 494 & 481.5564 & 0.112  & 552 & 484.9199 & -0.087 & 648 & 490.4859 & -0.437 \\
444 & 478.6549 & 0.251 & 495 & 481.6122 & 0.071  & 566 & 485.7333 & -0.109 & 649 & 490.5440 & -0.438 \\
445 & 478.7189 & 0.351 & 499 & 481.8423 & 0.026  & 567 & 485.7904 & -0.128 & 650 & 490.6030 & -0.424 \\
446 & 478.7747 & 0.309 & 500 & 481.9082 & 0.158  & 567 & 485.7921 & -0.098 & 664 & 491.4146 & -0.476 \\
447 & 478.8253 & 0.179 & 501 & 481.9544 & -0.048 & 569 & 485.9045 & -0.166 & 665 & 491.4719 & -0.493 \\
447 & 478.8274 & 0.215 & 504 & 482.1375 & 0.098  & 584 & 486.7694 & -0.303 & 666 & 491.5400 & -0.322 \\
457 & 479.4087 & 0.204 & 505 & 482.1974 & 0.127  & 585 & 486.8333 & -0.206 & 668 & 491.6425 & -0.560 \\
458 & 479.4702 & 0.262 & 509 & 482.4295 & 0.116  & 586 & 486.8901 & -0.229 & 681 & 492.3960 & -0.611 \\
460 & 479.5823 & 0.188 & 510 & 482.4819 & 0.017  & 587 & 486.9451 & -0.285 & 682 & 492.4595 & -0.520 \\
461 & 479.6400 & 0.180 & 511 & 482.5406 & 0.025  & 600 & 487.6953 & -0.392 & 699 & 493.4440 & -0.602 \\
464 & 479.8154 & 0.194 & 512 & 482.6006 & 0.056  & 601 & 487.7651 & -0.193 & 705 & 493.7895 & -0.665 \\
465 & 479.8748 & 0.214 & 515 & 482.7719 & -0.000 & 601 & 487.7622 & -0.243 & 706 & 493.8516 & -0.597 \\
466 & 479.9279 & 0.127 & 516 & 482.8296 & -0.008 & 603 & 487.8830 & -0.167 & 707 & 493.9032 & -0.711 \\
470 & 480.1643 & 0.188 & 516 & 482.8326 & 0.042  & 604 & 487.9327 & -0.312 & 722 & 494.7758 & -0.715 \\
471 & 480.2222 & 0.185 & 517 & 482.8892 & 0.015  & 612 & 488.3968 & -0.337 & 723 & 494.8371 & -0.662 \\
472 & 480.2755 & 0.100 & 518 & 482.9433 & -0.054 & 613 & 488.4523 & -0.383 & 724 & 494.8931 & -0.700 \\
474 & 480.3975 & 0.197 & 526 & 483.4092 & -0.047 & 614 & 488.5193 & -0.231 & 733 & 495.4140 & -0.747 \\
475 & 480.4534 & 0.157 & 527 & 483.4678 & -0.040 & 615 & 488.5730 & -0.308 & 739 & 495.7585 & -0.828 \\
476 & 480.5116 & 0.157 & 528 & 483.5244 & -0.068 & 618 & 488.7454 & -0.347 & 740 & 495.8167 & -0.828 \\
477 & 480.5736 & 0.224 & 529 & 483.5860 & -0.010 & 619 & 488.7985 & -0.434 & 785 & 498.4271 & -0.968 \\
478 & 480.6261 & 0.125 & 532 & 483.7593 & -0.032 & 620 & 488.8621 & -0.341 & 786 & 498.4876 & -0.929 \\
480 & 480.7420 & 0.117 & 533 & 483.8188 & -0.010 & 621 & 488.9207 & -0.333 &     &          &        \\
481 & 480.7976 & 0.072 & 534 & 483.8748 & -0.047 & 630 & 489.4422 & -0.372 &     &          &       \\
\toprule
\tabnotetext{*}{Individual errors in these timings are not explicitly included here.}
\tabnotetext{**}{E (cycle number).}
\tabnotetext{***}{Superhump maxima expressed as HJD - 2,456,000.}
\tabnotetext{****}{O - C value (in cycles) according to the ephemeris.}
\tabnotetext{}{T\_max (HJD) = 2,456,452.8035 + 0.058191 E.}
\end{tabular}  
\end{footnotesize}
\end{table*}

\begin{figure}
\centering
\includegraphics[angle=270,trim=1cm 4cm 2cm 3cm, clip, width=\columnwidth]{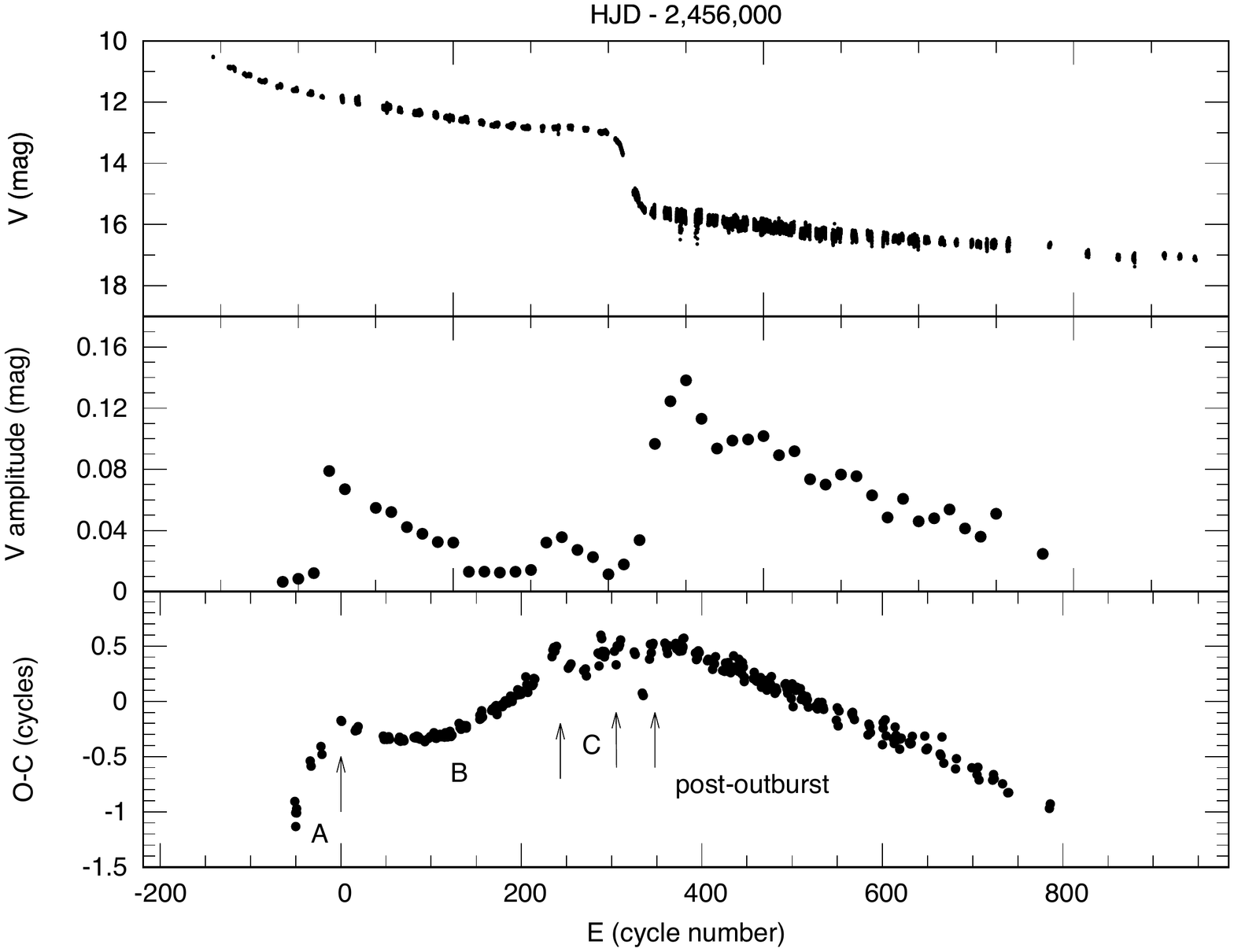}
\caption{
Photometric behaviour of \pnv\ during its superoutburst in 2013. Top: global light curve of  the observational campaign. Middle: amplitude variations of the superhumps along the superoutburst (see text). Bottom: O--C  diagram for the superhump maxima in the HJD 449.6--498.6 day interval with respect to the ephemeris given in Eq.~(\ref{ephem_all}). The arrows indicate the (approximate) location of transitions between different regimes of superhump period variations
(see text).}
\label{fig:fig-global}
\end{figure}
\bigskip


The top panel of Figure~\ref{fig:fig-global} displays the general photometric behaviour of the system during our campaign. Next, on the middle panel of Figure~\ref{fig:fig-global} is presented the variation of the amplitude of the superhump modulation, defined as the semi-amplitude of the sine wave that best fits the nightly photometric data.

The O--C residuals of the times of maximum light relative to the
ephemeris given by Eq. \ref{ephem_all} are shown in the lower panel of Figure~\ref{fig:fig-global}. The resulting O--C diagram is complex, but it displays a number of features that are usually observed in other SU~UMa-type systems \citep{kat15}. Among the most relevant features visible in this diagram we point out the following:

\begin{enumerate}


\item

During the first four days of the outburst a weak modulation (early superhumps) with a period $P_{\rm esh} \sim$ 0.057 d were visible in the light curve.

\item The onset of fully-grown ({\it stage A}  superhumps) took place in a short time-scale ($\sim$ 2 d) and involved an increase in the amplitude of the modulations. Their (mean) period, $P_{\rm sh(A)}\sim$ 0.059 d, was longer than $P_{\rm esh}$.

\item 

Once the superhump modulation reached full amplitude, the system entered {\it stage B} where the amplitude of the superhump decreased slowly, and the mean period became shorter ($P_{\rm sh(B)}\sim$ 0.058 d.) The upward curvature of the residuals during this stage (days HJD 449--466) signifies that the period of the superhumps was not constant, but increased over time. From a quadratic fit of the residuals in this interval, we find an increase rate of 
$dP_{\rm sh(B)}/dt= 5.8(4) \times 10^{-5}$.

\item
Before the end of the main eruption, the amplitude of the superhumps was found to grow larger ($\sim$ 0.10 mag). The system entered {\it stage C}, extending from day HJD 466 to the end of the main plateau (around day HJD 470), where the period of the superhump remained essentially constant ($P_{\rm sh(C)}\sim$ 0.0582 d).

\item
Worth noting is the increase of the amplitude variations as the system dropped about 3 mag between stage C and the post-outburst stage.

\item
After the end of the main eruption (day $\geq$ HJD 473) the superhumps were still visible with significantly larger amplitude $\sim$ 0.2 mag), and with a period which remains constant for at least the subsequent $\sim 25$ days. The period of the post--outburst modulation was shorter than $P_{\rm sh}$.

\end{enumerate}

 A summary of the main periodicities along the eruption is given
in Table~\ref{tab:tab_freqs}, as found in the next subsections.

\begin{figure}
\includegraphics[angle=270,trim=7cm 3cm 1cm 2cm, clip, width=\columnwidth]{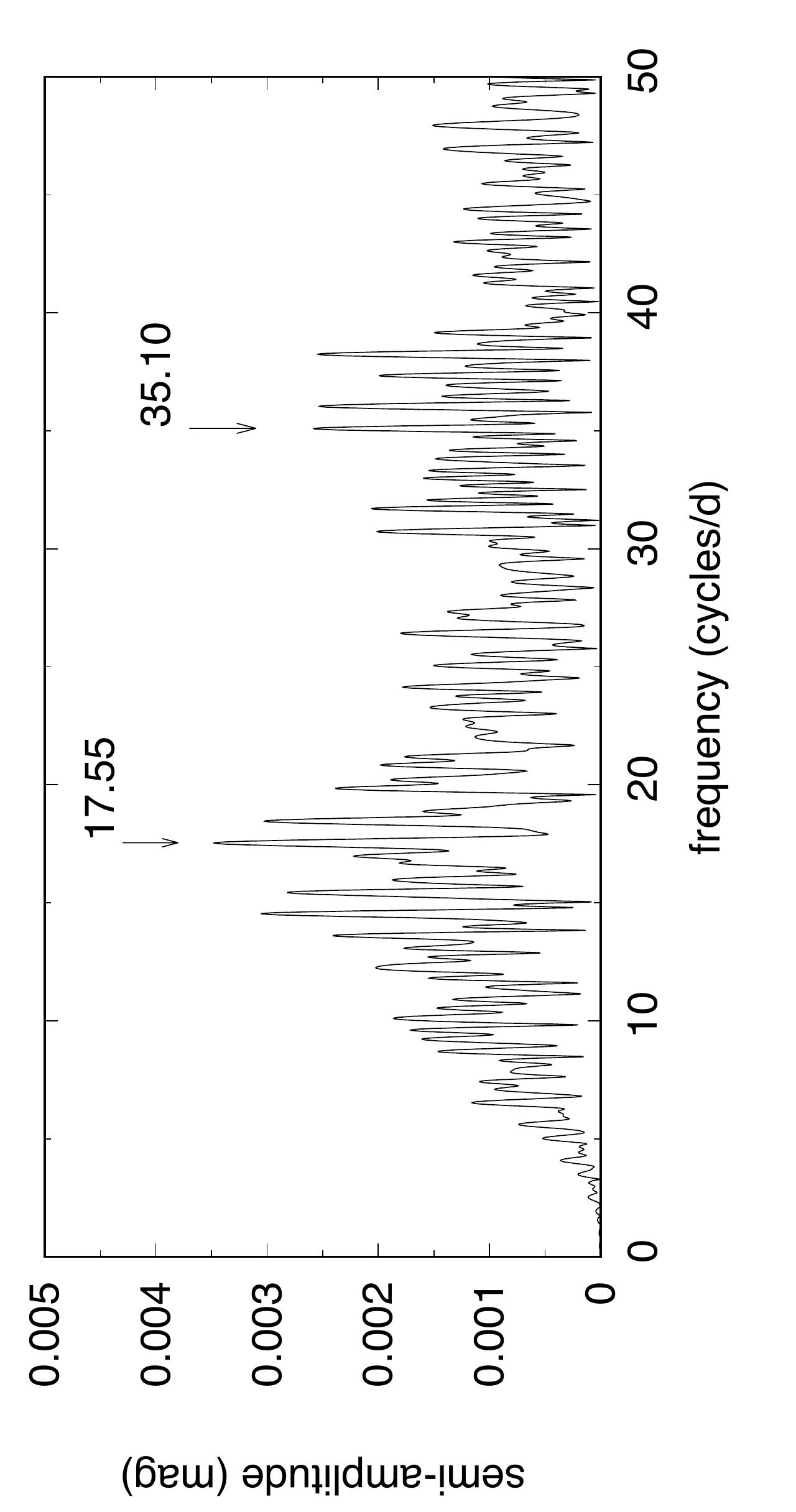}
\includegraphics[angle=0, width=\columnwidth]{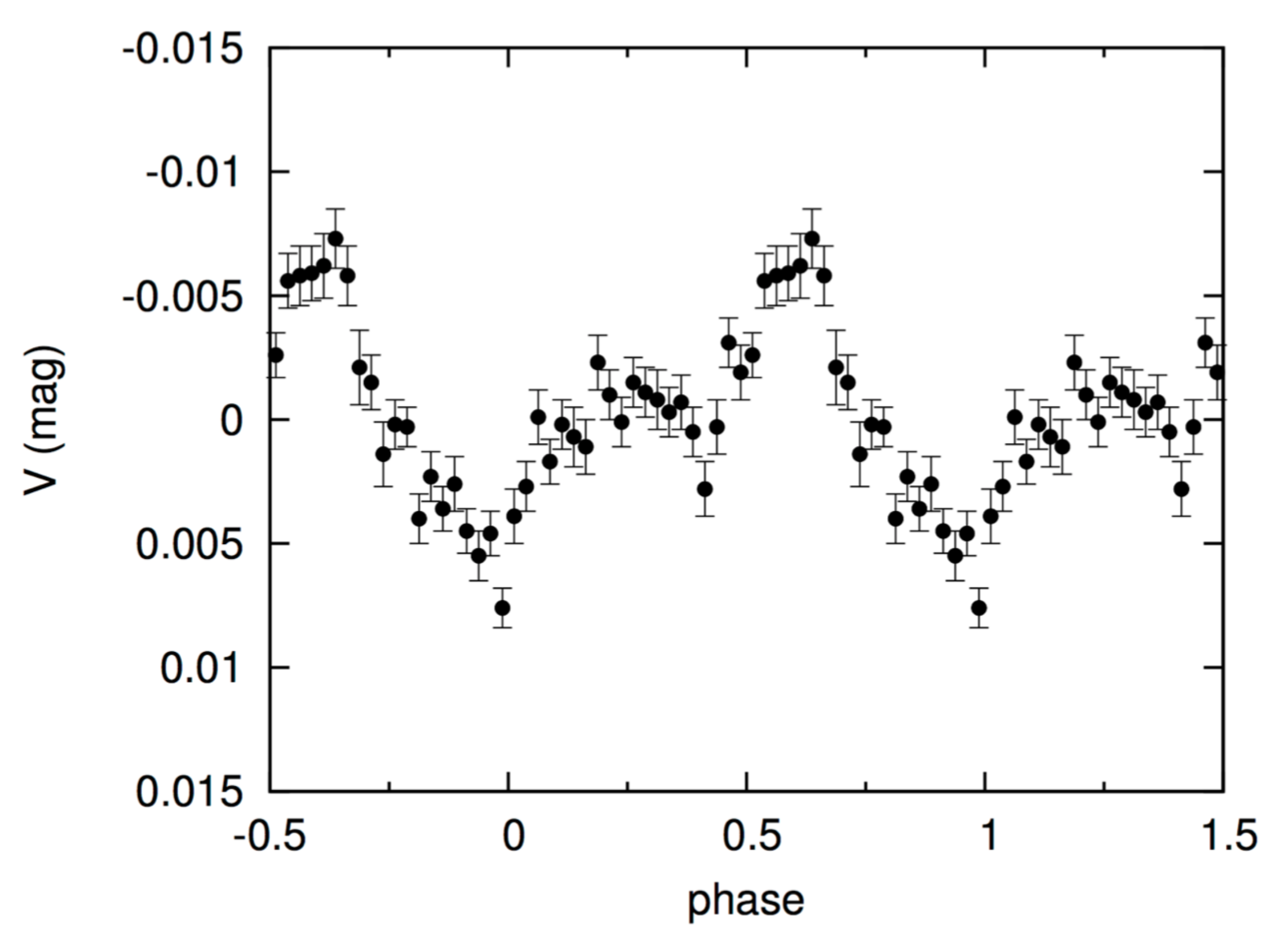}
\caption{
{\it Upper frame}:
Power spectrum during days HJD 445--448 (early superhumps), showing broad peaks centred at 17.55 cycles d$^{-1}$  and its first harmonic. {\it Lower frame}: Waveform of the early superhumps obtained after folding the data with $P_{\rm esh} = 0.05698$ d. The zero phase is arbitrary.
}
\label{fig:fig2_cba}
\end{figure}

\subsubsection{Early superhumps}
\label{earlysh}
From the beginning of our campaign, a weak modulation of about 0.010 mag full amplitude was observed in the light curves. This signal persisted over days HJD 445--448. The power spectrum of the spliced light curve covering this 4-day segment is shown in the upper frame of Figure~\ref{fig:fig2_cba}. It is dominated by two broad peaks centred at frequencies 35.10(3) and 17.55(3) cycles d$^{-1}$. These signals were weak, with amplitudes of 0.0036 and 0.0030 mag, respectively. Although they are barely detected above the noise, we interpret them as a likely manifestation of early superhumps \citep{kato14}. 

This photometric feature is known to be typical of WZ~Sge-type stars, and is not shown by any other type of dwarf nova. Although its physical origin is still under debate, there is increasing observational evidence that its period ($P_{\rm esh}$) is essentially equal to $P_{\rm orb}$ \citep{pea96,kat15}. A folded curve of the spliced light curve with $P_{\rm esh} = 0.05698(9)$ d is also shown in the lower frame of Figure~\ref{fig:fig2_cba}, which shows the double-humped pattern characteristic of early superhumps. The value of $P_{\rm esh}$ obtained in this paper is slightly different from, but consistent with, 
the value of 0.05706(2) d reported in \citet{kato14}.

\begin{figure}
\includegraphics[angle=-90,trim=7cm 3cm 1cm 2cm, clip,width=\columnwidth]{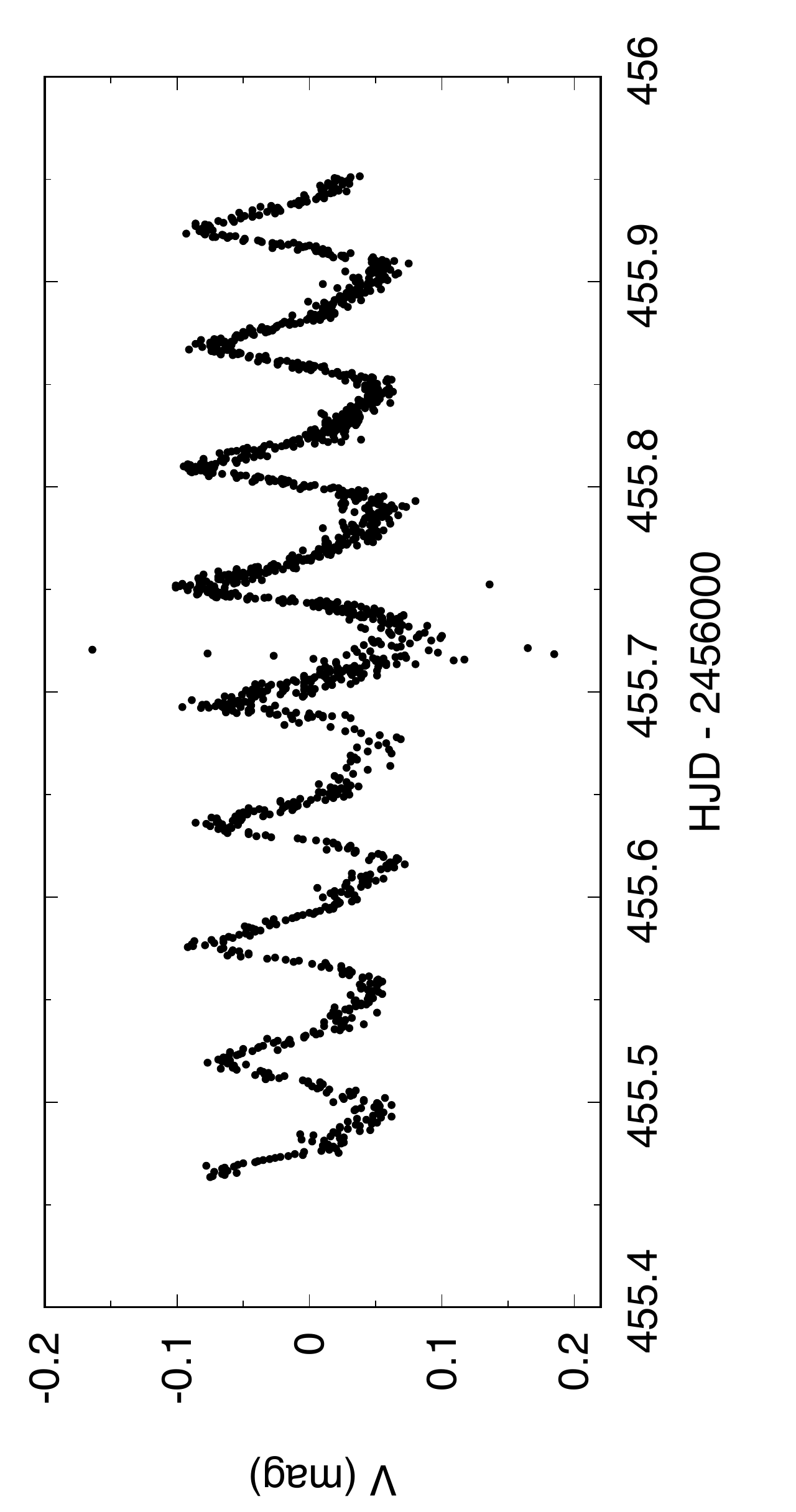}
\includegraphics[angle=-90,trim=7cm 3cm 1cm 2cm, clip,width=\columnwidth]{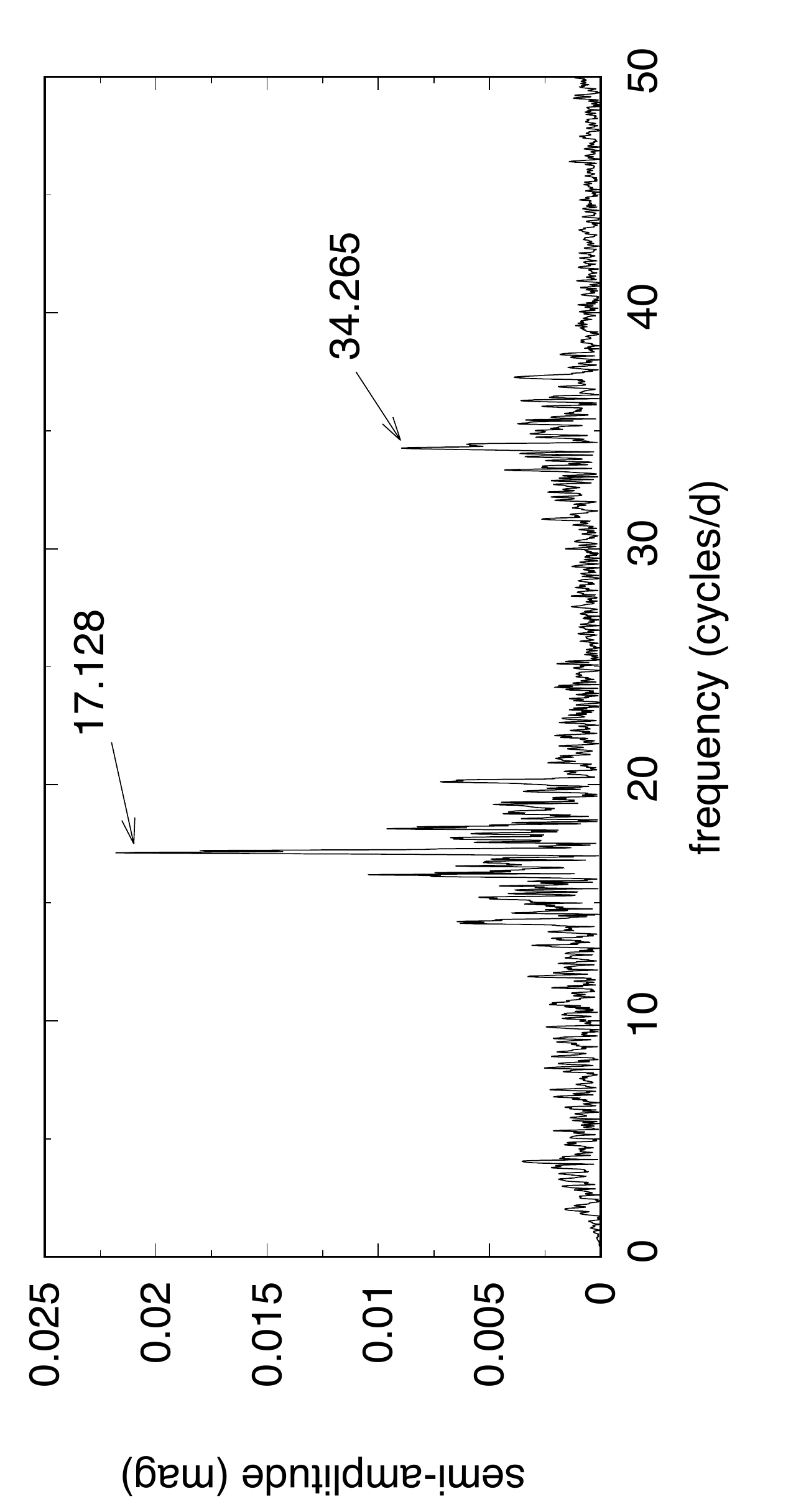}
\includegraphics[angle=-90,width=\columnwidth]{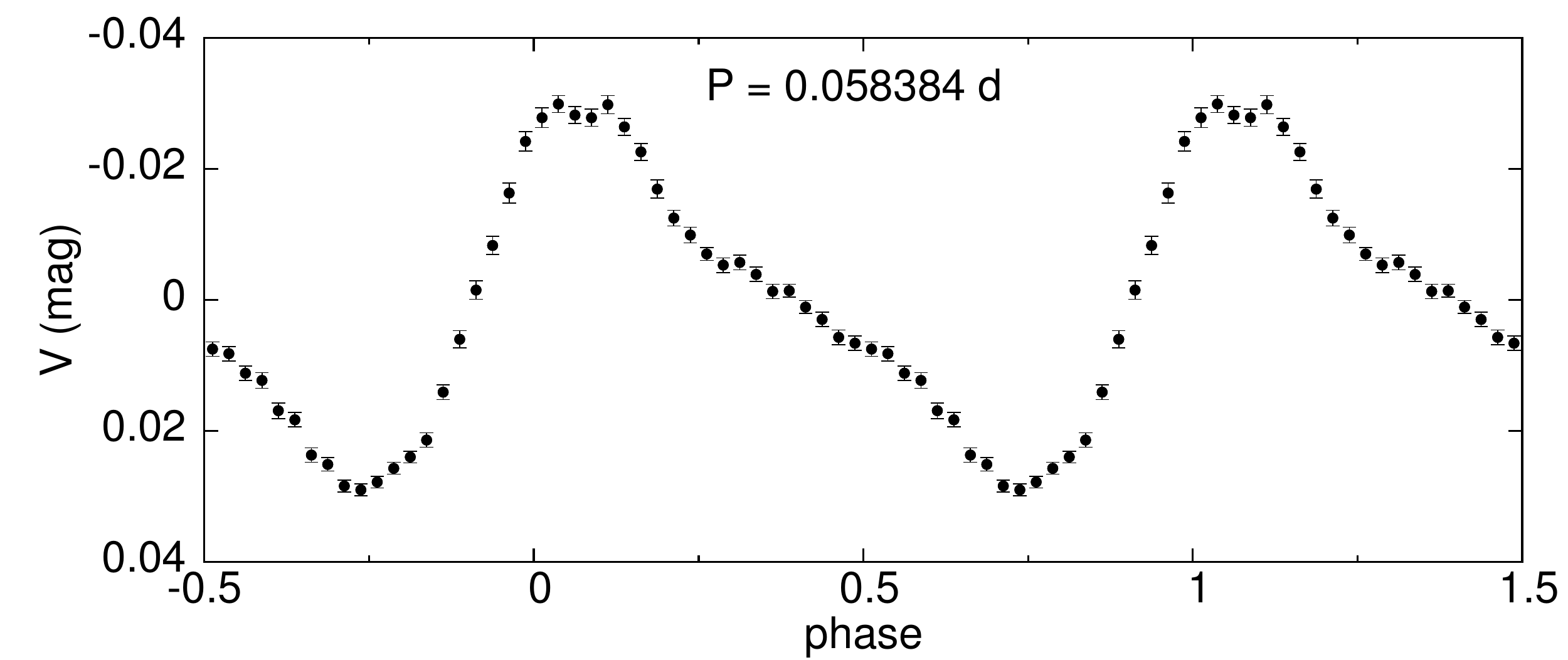}
\caption{
{\it Upper frame}:
A 12--hour spliced light curve obtained on day HJD 455, dominated by large-amplitude common superhumps. The zero level in the figure corresponds to $V \approx 12.1$ mag. {\it Middle frame}: Power spectrum during the common superhump era (days HJD 442--466), with main peaks centred at frequencies 17.128 and 34.265 cycles d$^{-1}$. {\it Lower frame}: Mean waveform of the common superhump obtained after folding the data on $P=0.058384$ d. The zero phase is arbitrary.
}
\label{fig:fig3_cba}
\end{figure}

\subsubsection{Common superhumps}

 The double-humped pattern of the early superhumps turned into single-peaked humps on day HJD 449. Over days HJD 449--451, the mean amplitude was around $\sim0.007$ mag and the period was $\sim$4\% longer than the period found for early superhumps. The dominant signal occurred at 16.83(8) cycles d$^{-1}$ corresponding to a period of 0.0594(3)~d. Since the modulation was better defined in this 2-day interval, we were able to determine the times of maximum in the signal. We identified a total of 9 maxima in the HJD 449.6--451.6 day interval, and obtained a period of 0.05934(11)~d (corresponding to a frequency of 16.85(3) cycles d$^{-1}$) from a linear regression. This value is fully consistent with the one found from the Fourier analysis. The modulation over this 2-day segment is interpreted as {\it stage}-A superhumps. The period we find is close to, but slightly different from, the value of 0.05883(6)~d reported in \citet{kato14}.

Fully-grown, large-amplitude superhumps were finally observed on day HJD 452 (amplitude of 0.10 mag). As a representative example, we show in the upper frame of Figure~\ref{fig:fig3_cba} the light curve from day HJD 455. As the eruption proceeded, the mean amplitude of the superhumps decreased (as shown in the middle panel of Figure~\ref{fig:fig3_cba}). The variation in amplitude was smooth in the HJD 452--466 day interval. We formed a spliced light curve in this interval, and obtained the power spectrum shown in the middle frame of Figure~\ref{fig:fig3_cba}. The strongest signals occurred at $f_1 =$ 17.128(3)  and $f_2 =$ 34.265(3) cycles d$^{-1}$. 
They are interpreted as the frequency of {\it stage-B}
superhumps (period of $P_{\rm sh}=0.058384(10)$ d) 
and its first harmonic, respectively.
Other (weaker) peaks, not shown in Figure~\ref{fig:fig3_cba}, are found at $f_3 =$ 51.381(6) and $f_4 =$ 68.403(6) cycles~d$^{-1}$. The mean waveform of the superhump modulation during this interval is shown in 
the lower panel of Figure~\ref{fig:fig3_cba}.

We note that after subtracting the superhump signal and its harmonics, the power spectrum of the residual light curve showed peaks at 17.20 and 17.28 cycles d$^{-1}$. But we do not give any physical significance to these detections, and interpret them as the result of period and amplitude variations of the superhump wave during the eruption.

The amplitude of the superhumps increased around day HJD 466, and decreased thereafter until the end of the main eruption (stage C). The strongest signal in the power spectrum in the HJD 466--470 day interval occurred at 17.191(10) cycles d$^{-1}$, corresponding to a period of $0.05817(3)$ d, with additional peaks at higher harmonics.

\begin{figure}
\includegraphics[angle=-90,trim=7cm 3cm 1cm 2cm, clip,width=\columnwidth]{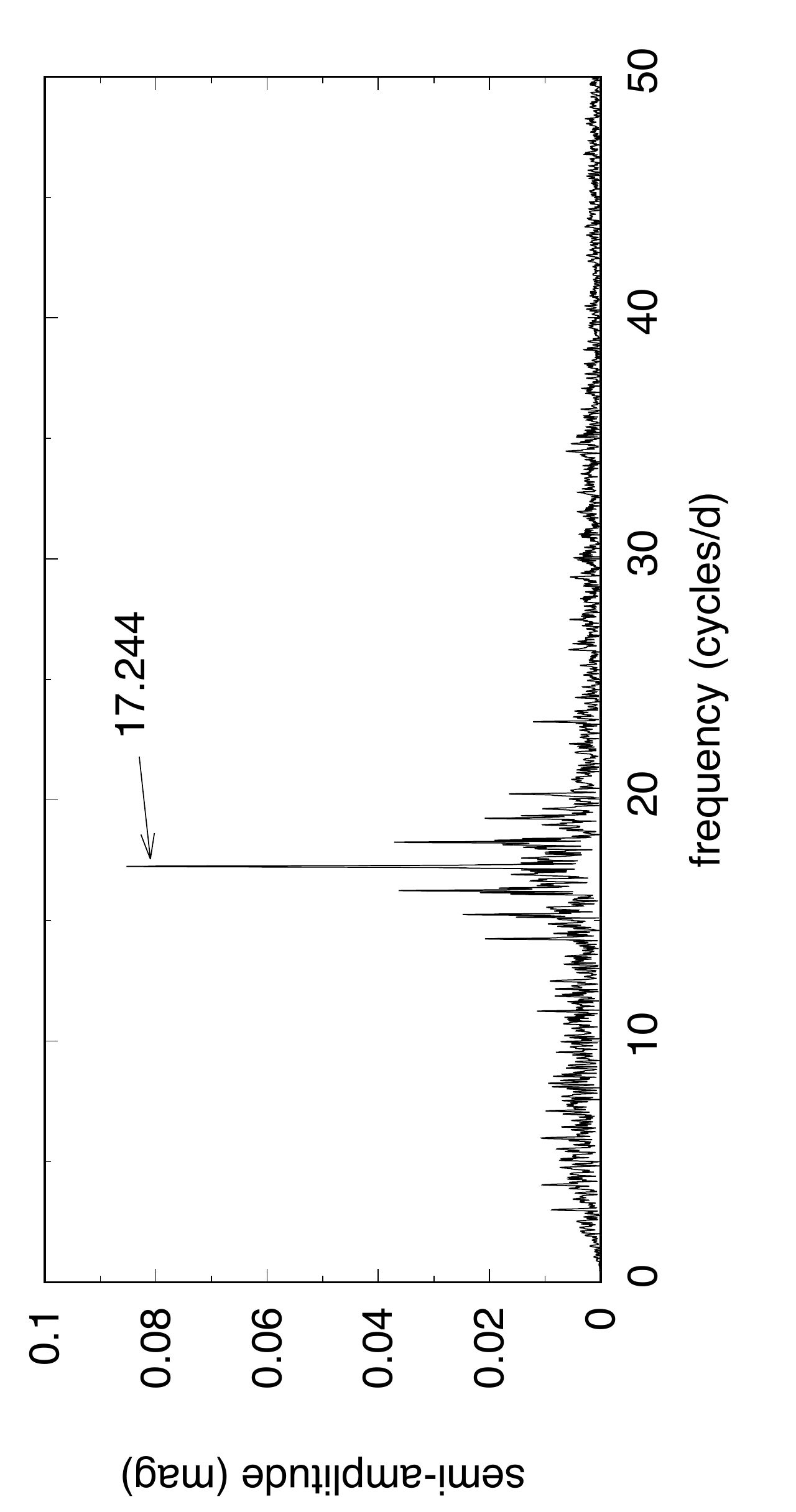}
\includegraphics[angle=-90,width=\columnwidth]{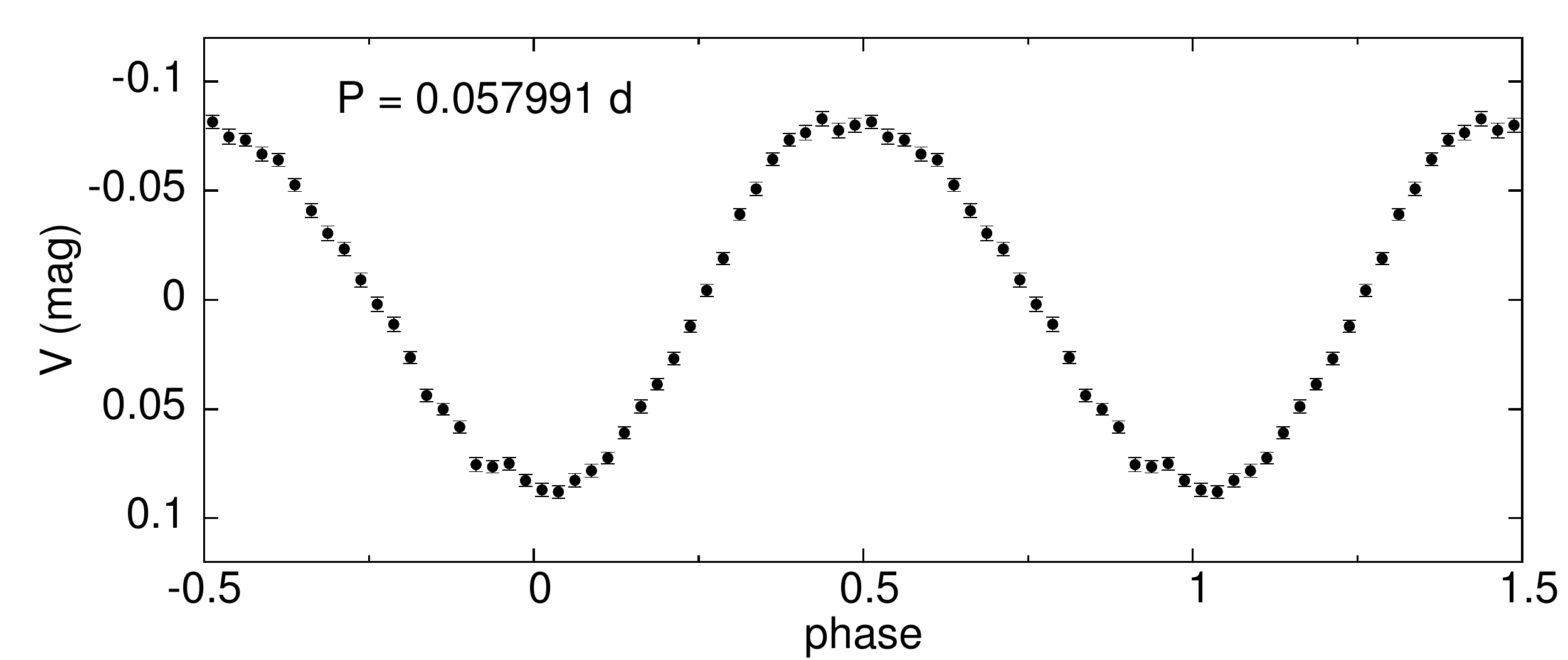}
\caption{
{\it Upper frame}:
Power spectrum after the main eruption 
(day HJD $\geq$ 473) showing a strong peak centred at 
17.244(1) cycles d$^{-1}$ (post--outburst superhump).
{\it Lower frame}: 
Mean waveform of the post--outburst superhump, obtained after folding the data
on $P = 0.057991$ d. The zero phase is arbitrary.
}
\label{fig:fig4_cba}
\end{figure}

\subsubsection{Post--outburst stage}

Once the main eruption was over, the light curve was still dominated by superhumps, but now with significantly larger amplitude ($\sim 0.15$ mag) which decreased slowly ($\sim 0.013$ mag d$^{-1}$). This behaviour remained essentially unchanged for nearly 20 days of our observations after the main fading.

We formed a spliced light curve including all the observations from  day HJD $\geq 473$, and found a power spectrum (shown in the upper frame of Figure~\ref{fig:fig4_cba}), with a peak at 17.244(1) cycles d$^{-1}$. This is interpreted as the frequency of the post-outburst superhump, and dominates the spectrum. Higher-order harmonics were also found, but their amplitude was very low ($< 0.0065$ mag). This signifies that the waveform of the post-outburst superhump  was nearly sinusoidal. The lower frame of Figure~\ref{fig:fig4_cba} shows that this is indeed the case.

\begin{table*}
 \centering
\caption{
Mean photometric periods and frequencies of the superhump modulation during the 2013 superoutburst of \pnv. The errors in parenthesis correspond to the last significant figures. 
}
\centering
\begin{tabular}{llll}
Photometric Periods\\
\hline
\multicolumn{1}{c} {Time interval}  &
\multicolumn{1}{c} {Period}      &
\multicolumn{1}{c} {Frequency}         &
\multicolumn{1}{c} {Comments}        \\
\multicolumn{1}{c} {(HJD--2456000)}  &
\multicolumn{1}{c} {(d)}         & 
\multicolumn{1}{c} {(cycles d$^{-1}$)}  &  
          \\
\hline
445--448 & 0.05698(9)   & 17.55(3)   & early superhumps \\
449--451 & 0.0594(3)   &  16.83(8)   & {\it stage} A \\
452--466 & 0.058384(10)  & 17.128(3) & {\it stage} B \\
466--470 & 0.05817(3)  &  17.191(10)  & {\it stage} C \\
473--498 & 0.05799(1)  &  17.244(1)  & post outburst \\
449--498 & 0.0581910(1)  &  17.1847(3)  & mean superhump period\\
\hline
\label{tab:tab_freqs}
\end{tabular}
\end{table*}

\subsection{Photometry near and at minimum light}
\label{sec:min-light}

\begin{figure}
\includegraphics[width=1.05\columnwidth]{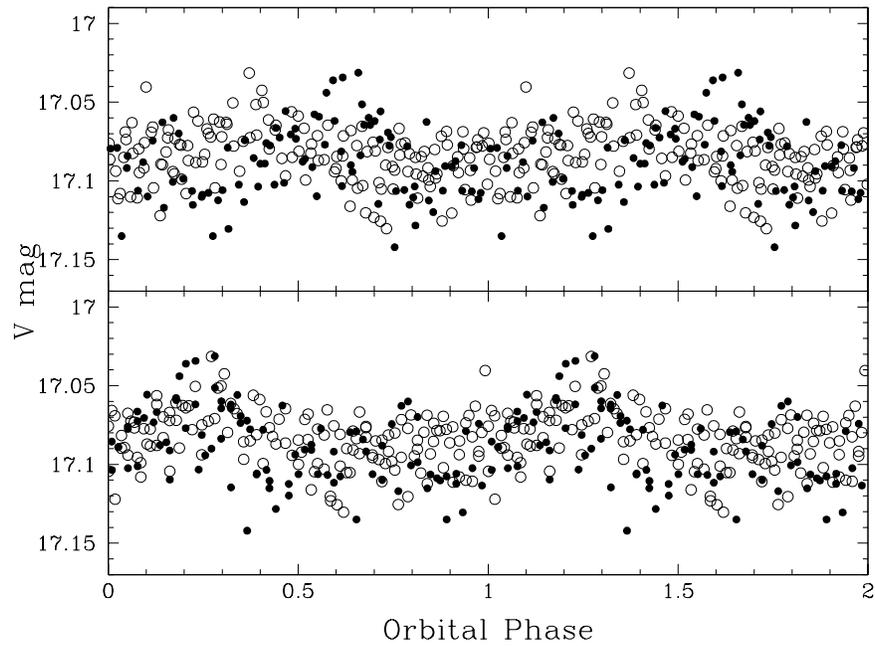}
\caption{Upper: $V$ light curves obtained near minimum light in 2013. Open dots are from September 2 and 
filled dots from September 25. The size of the points corresponds to the mean individual errors. The orbital phases have been taken from the ephemeris obtained in this paper. Lower: The two nights folded with the ephemeris obtained by using the PDM technique (see more details in text).
}
\label{fig:fot0225}
\end{figure}

\begin{figure}
\includegraphics[width=1.05\columnwidth]{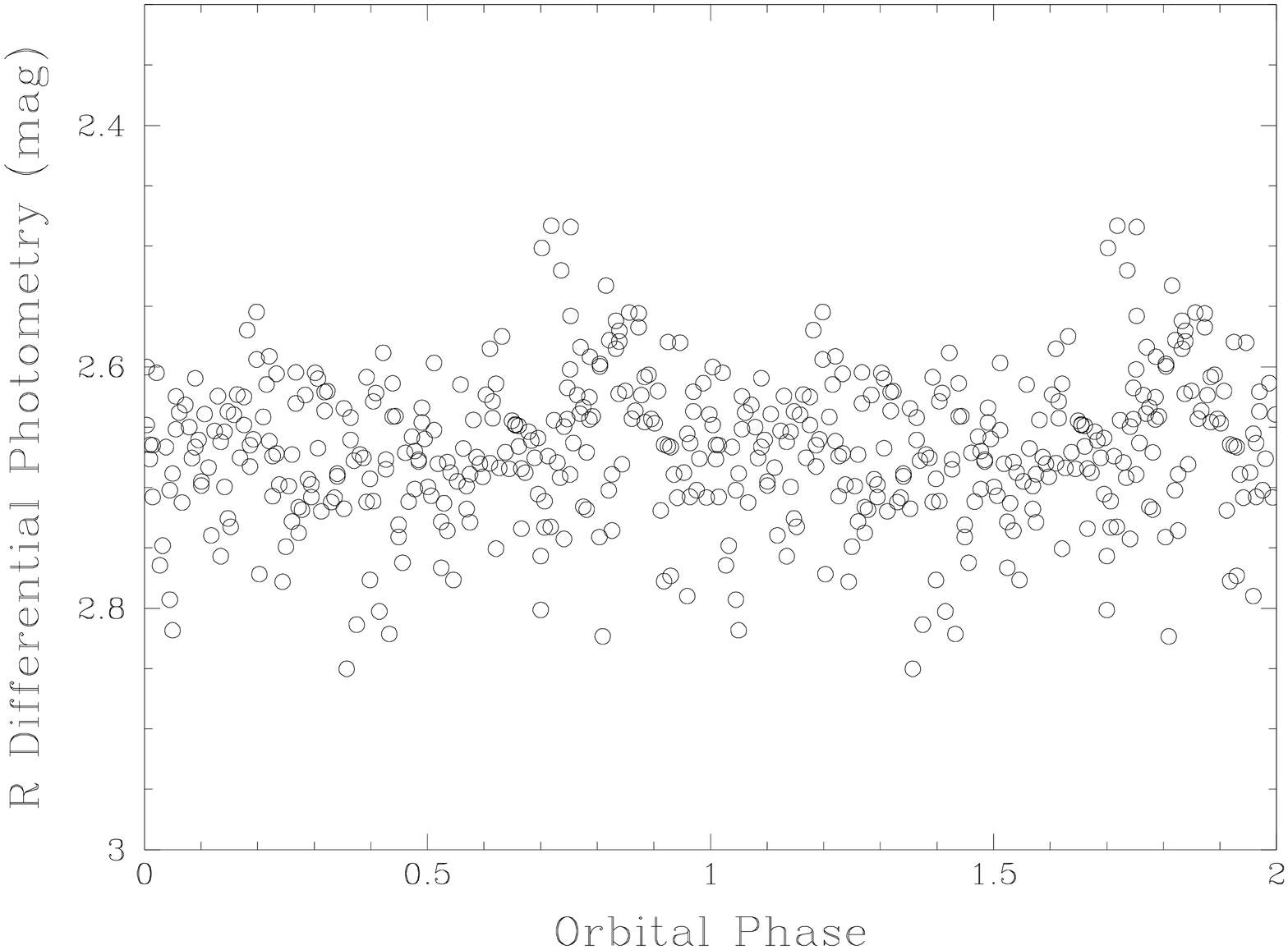}
\caption{Differential $R$ light curve obtained at minimum light in July 18, 2018.  The size of the individual errors ($\sim 0.03$ mag) is rather large. The orbital phases have been taken from the ephemeris obtained in this paper. No obvious orbital modulation, within the errors, is detected at this level.
}
\label{fig:fotR2018}
\end{figure}

As detailed in Section~\ref{sec:observations}, we took two runs near minimum light covering around one orbital period each. The light curves are shown in the upper panel of Figure~\ref{fig:fot0225}. When folded with the orbital period (0.05698 d), the light curves seem to
be out of phase. But this is not surprising: over the 23 days (nearly 400 orbital cycles) elapsed between both runs, an uncertainty of 0.0001 days in $P_{\rm orb}$ involves an 
uncertainty of 0.7 in phase.
We carried out a period analysis of both nights using the Phase Dispersion Minimization (PDM) technique \citep{ste78} in the {\it Peranso} package \citep{pau16}. This technique
is frequently used to detect variations of superhumps in SU UMa systems \citep[e.g.][]{kato14}. The lower frame in Figure~\ref{fig:fot0225} shows the results of combining the two nights with the best period estimate (0.0576 d) determined from the PDM technique. Assuming that the observed light comes from the accretion disc, the zero point obtained in this case is HJD 2456537.6946 (time of inferior conjunction of the secondary). The period found using the PDM method yields a value which is still close to the post-outburst state, but the sinusoidal shape is gone. There is only a small peak around phase 0.25. No double modulation with orbital period is found as would be expected in a bounce-back object. Further observations in the $R$ band were obtained in 2018, July 18 covering three orbital cycles. Since the night was not photometric, we were unable to make absolute calibrations and only differential photometry is shown in Figure~\ref{fig:fotR2018}. No obvious orbital modulation was detected within the individual errors, which are rather large ($\sim 0.03$ mag).

\section{Discussion}
\label{sec:discus}

The values derived for the superhump period in {\it stages} A and B allowed us to estimate the mass ratio of the system through the known ``superhump excess'' $\epsilon-q$ relations \citep{pat98,pat11,kat13}. Considering the lack of a reliable determination
of the orbital period from spectroscopic observations, we assume here that $P_{\rm orb}$
is equal to the period of early superhumps.
We find $\epsilon_A=0.042(5)$ and $\epsilon_B=0.024(2)$. Thus, our estimates for the mass ratio are $q_A=0.12(2)$ and $q_B=0.10(1)$, respectively.
Comparing these two values with the $\epsilon-q$ relation shown in \citet[][Figure 19 ]{bea17} we can see clearly that $\epsilon_B$ is well within the expected value, while $\epsilon_A$ is not. This is further supported by using the updated \citet{sas84} relation in \citet[][Eq. 4]{otu16}, which for our assumed orbital period gives $\epsilon = 0.019(10)$. Although we are inclined to use the {\it stage B} results, we point out that both values are suggestive of a low-mass donor, although as pointed out in Section~\ref{sec:intro} \citep[e.g.][see their Figure~7]{otu16}, empirical relations $\epsilon$-$q$ at low-$q$ values may carry large systematic uncertainties. 
For a typical white dwarf with mass $\sim0.8$ M$_{\odot}$ \citep{zor11} and mass ratio $q\sim0.1$, the mass of the secondary is very close to the sub-stellar limit i.e. $0.072 M_{\odot}$ \citep{cha97}. Further characterisation of systems like \pnv\ will allow us to discern empirically where this limit lies for mass-losing donors.

It has been noted by many authors, both theoretically and observationally, that the CV orbital period distribution should present a sharp cut-off at about $\sim80$ min, usually termed as the minimum period \citep[e.g.][]{rap82,rak98,gan09}.
\pnv\ has an orbital period of about 82 min, very close to the minimum period, which makes it difficult to discern whether it is approaching to or receding from this minimum orbital period. Before asserting its true nature, we could look at some observational features in those CVs systems around the minimum orbital period. Most of these systems possess WZ Sge-like features. 
Their optical spectra are mostly dominated by the white dwarf and accretion disc itself, with no visible features from the donor. Since the rate of accretion is an order of magnitude smaller than that for systems before reaching the minimum period ($\dot{m}\sim10^{-11}$ M$_{\odot}$ yr$^{-1}$), the accretion discs become very faint, and the broad absorption lines of the white dwarf become visible below $\sim5000$ \AA) e.g. WZ Sge \citep{hea08}. On the other hand, the donor's observed properties should vary significantly for systems with the same orbital period but evolving towards or away the period minimum. This is a consequence of the donor's temperature steep relationship as a function of orbital period \citep[e.g.][]{kni11}.
From the superhump analysis presented here, we suspect that the system is approaching the period minimum and thus, the donor is probably a late-M dwarf spectral type with an observed effective temperature of $\sim2400$ K \citep{kni11}.

The donor of \pnv\ is therefore an ideal candidate for NIR time-resolved spectroscopy \citep[e.g. SDSS~J143317.78+101123.3,][]{her16}, which would render a fully independent measurement of the orbital period and the mass ratio, to confirm or reject its sub-stellar nature. This is particularly important since few low-q systems have been observed in outburst and which are also capable of dynamical measurements of their components \citep[Figure~3 in ][]{kat13}. Thus, \pnv\@ could be a system to calibrate the empirical superhump relations in this poorly-explored region of parameter space. 

\section{Conclusions}
\label{sec:conclusions}

We have presented a long-term study of the 2013 superoutburst of \pnv\ from its peak to quiescence. Our main results are as follows:
\begin{itemize}

\item The observed early superhumps suggest an orbital period of $P_{\rm orb}=0.05698(9)$ d, which locates \pnv\@ close to the minimum of the
orbital period distribution in CVs.

\item From {\it stages} A and B and early superhump periods, we found the mass ratio to be $q_A=0.12(2)$ and $q_B=0.10(1)$, respectively.

\item Based on the obtained values of the mass ratio, we claim that the donor in V1838 Aql is a low-mass star rather than a sub-stellar object, and the system is approaching the period minimum.

\end{itemize}

Given the long interval between outbursts in low-q systems, it is of paramount importance to confirm by dynamical methods the orbital parameters of such systems. This would indicate which systems may be used in order to calibrate the empirical superhump excess relations.

\section*{Acknowledgements}

The authors are indebted to DGAPA (Universidad Nacional Aut\'onoma de M\'exico) support, PAPIIT projects IN111713, IN122409, IN100617, IN102517, IN102617, IN108316 and IN114917. GT acknowledges CONACyT grant 166376. JE acknowledges support from a LKBF travel grant to visit the API at UvA. JVHS is supported by a Vidi grant awarded to N. Degenaar by the Netherlands Organization for Scientific Research (NWO) and acknowledges travel support from DGAPA/UNAM. E. de la F. wishes to thank CGCI-UdeG staff for mobility support. We thank the day and night-time support staff at the OAN-SPM for facilitating and helping obtain our observations.

\end{document}